# Augmented Collective Intelligence in Collaborative Ideation: Agenda and Challenges

Emily Dardaman

BCG Henderson Institute, dardaman.emily@bcg.com

Abhishek Gupta

BCG Henderson Institute, gupta.abhishek@bcg.com

AI systems may be better thought of as peers than as tools. This paper explores applications of augmented collective intelligence (ACI) beneficial to collaborative ideation. Design considerations are offered for an experiment that evaluates the performance of hybrid human–AI collectives. The investigation described combines humans and large language models (LLMs) to ideate on increasingly complex topics. A promising real-time collection tool called Polis is examined to facilitate ACI, including case studies from citizen engagement projects in Taiwan and Bowling Green, Kentucky. The authors discuss three challenges to consider when designing an ACI experiment: topic selection, participant selection, and evaluation of results. The paper concludes that researchers should address these challenges to conduct empirical studies of ACI in collaborative ideation.

**CCS CONCEPTS** •Human-centered computing ~Collaborative and social computing ~Empirical studies in collaborative and social computing •Human-centered computing ~Interaction design ~Interaction design process and methods ~Activity centered design

## 1 INTRODUCTION

Understanding the future of collaborative ideation requires understanding collective intelligence (CI): how to define, design for, and measure it. Put simply, CI is the ability of groups to solve problems more effectively than individual members could on their own [1]. Past studies of CI have discovered it in ants [2], animals [3], humans [4], and artificial intelligence (AI) [5]: across many compositions of groups, CI emerged. The last twenty years have seen studies shift from individual performance, often measured in IQ, to collaborative performance as CI[6]. Studies of human CI have revealed that it is not just a function of the average group member's IQ or even its most intelligent member; it is mediated by the group's size, social perceptiveness, and communication ability [4]. Recent increases in AI have opened new possibilities for CI, as seen in the newly created subfield of augmented collective intelligence (ACI). ACI researchers, including the authors, seek to understand factors driving performance in hybrid human-AI collectives. Historically, AI systems have been treated as tools for a good reason: their capacities were limited, especially in creative applications requiring cross-disciplinary thinking. Their unreliability made them poor choices to rely on for many uses. However, as capabilities increase, the delegation of roles between humans and advanced AI systems, such as LLMs, is shifting. Mapping roles in a real-time ideation can reveal places where humans are outperformed or on equal footing to LLMs.

## 2 STUDYING MIXED PERFORMANCE

Existing literature from nearby fields like CI or human-computer interaction (HCI) frequently covers 1:1 human-AI interactions, i.e., one human working with one AI system [7] or groups made of all humans [8] or groups composed of just AIs [9]. Still, there needs to be more research on mixed-team performance. We seek to remedy that by designing an experiment where mixed composition collectives of humans and advanced AI systems, such as LLMs, participate in collaborative ideation. This experiment will use Polis [21], a real-time response collection tool that prompts participants with questions, solicits responses from each and encourages everyone to upvote their favorite ideas. It maps opinion-based groups and axes of disagreement in real-time and allows group members to find common ground.

### 2.1 Case study: Taiwan

Polis's design facilitates a natural process of divergence and convergence that drives inclusion, creativity, and converging on the solutions; it is beneficial for complex or divisive topics that stall in other ideation formats. In Taiwan, a team of "civic hackers" used Polis to develop citizen consensus on contentious policy issues, including how to regulate Uber [10] and whether or not to change time zones [11]. Using Polis revealed the common concerns at the heart of these debates (safety and independence from China, respectively). It provided direct inputs to policies developed by Taiwan's digital minister, Audrey Tang.

### 2.2 Case study: Bowling Green

Bowling Green, Kentucky, also used Polis to generate ideas for improving residents' experience [12]. Despite over 2,000 people participating in the ideation, the results showed overwhelming consensus on most issues and clarified core disagreements on the remainder. Popular ideas stood out among the data and were then usable by city officials.

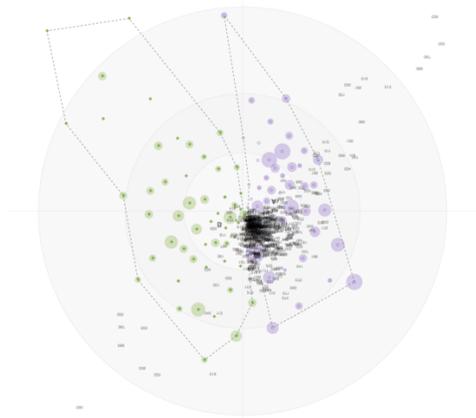

Figure 1: Polis response data revealed two clusters, elegantly simplifying group dynamics
to participants' position on a core disagreement. [12]

## 3  STUDY DESIGN; THREE CHALLENGES

Polis is a promising tool to measure the effectiveness of human-AI collectives. Designing an effective study includes three challenging components: topic selection, participant selection, and evaluation of results.

### 3.1  Challenge #1: Topic Selection

For topic selection, we are designing a set that escalates in complexity and difficulty to establish a baseline and identify performance limits. Participants take on roles such as suggester, synthesizer, or fact-checker during ideations. As the topics grow in complexity, these roles may shift. Before the first topic, we will develop hypotheses about which roles might be taken by AI or human participants. Then, as part of the measurement process, we will map the changing roles on a matrix. We will focus on topics most relevant to the lower end of the complexity spectrum, which are amenable to our study. Characteristics of these questions are a sufficiently narrow subject (e.g., planning for a software development sprint), ease of judgment by subject matter experts (e.g., grading the rigor of a mathematical proof), and the ability to converge on a single solution through consensus (e.g., picking a book for a reading group). On the higher end of the complexity spectrum, topics will likely have broad focus areas (e.g., company business strategy, political future of a country), which are more difficult to judge. There might be many different ways to solve the same problem, or it might be difficult to gain consensus. Microservices architecture in a software infrastructure or fashion trends would be poor topic choices. Deliberately choosing topics gives the experimental approach a greater chance of success.

### 3.2 Challenge #2: Group Selection

Choosing which individuals to include and how to include them in the collective is even more complex. In addition to the usual concerns about finding available, willing participants (who are as close to a representative group as possible), we must consider how group configurations are likely to influence ACI. Human demographic diversity, for example, affects group performance in different ways. On average, women have a higher social perception, so they are likely to improve CI; moderate levels of demographic diversity improve attentiveness, but too much impairs cooperation, and age diversity can reduce CI by introducing hierarchy-type dynamics into the group [4]. AI system selection will also impact ACI results. Each AI system has distinct capabilities, affecting the roles it can play in ideation and the quality of its interactions with group members. It is critical to avoid naively extrapolating results from human-ChatGPT [13] teams to teams featuring Claude [14], Sparrow [15], or future models. Combining models in mixed teams might create interesting results, although models have much broader differences in ability than people, so the utility of weaker models may be limited. In conducting this research, we do not expect to find a fixed dynamic where machines provide X, and only humans can provide Y. Research modeling human contribution must avoid falling into the trap of underestimating capability advances. With many features of general intelligence forecast to arrive within the decade [16], we are less interested in defining an "ownable" human role in hybrid teams and more interested in noting how such teams can be effective now and what direction they are moving in.

In particular, the configurations will change depending on the type of topic at hand, and a systematic study of ACI can help surface that. Other group selection choices that might result in new findings include varying the human/AI ratios, changing the group size, and disclosure/non-disclosure of the AI systems. AI systems may present relatively homogenous ideas or share similar blindspots, so having multiple instances of the same model running may not provide additional value. Larger groups may demonstrate higher ACI since CI is an emergent property. Finally, experimenting with disclosing or not disclosing the presence of chatbots to the humans in the ideation can be used to test for automation bias [17] and algorithm aversion bias [18] and how those impact ACI.

### 3.3 Challenge #3: Evaluation

Just as much work will go into building IQ tests to measure individual intelligence, we expect designing the testing mechanism for joint human-AI collectives in collaborative ideation to be difficult and require iteration; this is a common struggle in CI and extends to ACI as well. A simple baseline is to use subject matter experts to evaluate the outputs from a collective (which would serve the lower end of the spectrum for topic complexity). For topics on the higher end of the complexity spectrum, we expect that novel evaluation mechanisms and associated measures will have to be developed. These evaluation mechanisms can borrow from work in metrology studies conducted at the National Institute of Standards and Technology (NIST) [19] and the *Measuring Collective Intelligence* project at the MIT Center for Collective Intelligence (CCI) [20]. We also foresee a vital role in the assessment offered by each participant of all other participants in their overall contributions to finding the solutions as a collective towards the problems it is trying to solve. Assessment helps identify whether introducing AI agents in the human-AI collectives yields additional value through direct contribution or assisting human agents.

## 4   CONCLUSION

As AI systems improve and become more widely available, collaborative ideation can include AI agents as peers, not tools. When deployed safely, ACI can be a powerful force to help organizations gain a competitive edge in the market. Building a study to measure ACI's role in collaborative ideation requires designers to face three challenges. First is choosing the topics most germane to ACI's emergent capabilities. Next is selecting the composition of each human-AI collective, and the final is evaluating the collective's outputs during the ideation. This agenda and set of challenges invite participation from researchers and practitioners to chip away at the problems to help realize the potential of ACI.

## ACKNOWLEDGMENTS

The BCG Henderson Institute funded and supported this work under Abhishek Gupta's Fellowship on Augmented Collective Intelligence.